\documentclass[nofootinbib,eqsecnum,prd,aps,epsf]{revtex4}
\usepackage{times}
\usepackage{amssymb}
\usepackage[]{graphicx}
\newcommand{\be}{\begin{equation}}
\newcommand{\ee}{\end{equation}}
\newcommand{\bea}{\begin{eqnarray}}
\newcommand{\eea}{\end{eqnarray}}
\begin{document}


\title[topological censorship and chronology protection]{Topological censorship and chronology protection}

\author{John L. Friedman$^1$ and Atsushi Higuchi$^2$}
\affiliation{$^1$Department of Physics, University of Wisconsin-Milwaukee,\\ 
			PO Box 413, Milwaukee, Wisconsin, US\\
$^2$Department of Mathematics, University of York,\\
Heslington, York, YO10 5DD, UK\\
}

\begin{abstract} Over the past two decades, substantial efforts have been made to understand 
the way in which physics enforces the ordinary topology and causal structure that we observe, from subnuclear to cosmological scales.  We review the status 
of topological censorship and the topology of event horizons; chronology protection in 
classical and semiclassical gravity; and related progress in establishing quantum energy 
inequalities. \\

\normalsize This article is dedicated to Rafael Sorkin, whose friendship and 
tutoring from third grade 
onward is responsible for one of us (JF) having spent his adult life in physics and 
whose work has inspired both of us.
 
\end{abstract}

\maketitle                
\setcounter{page}{1}

\section{Introduction}
In addition to the gravitational waves and black holes that reside in 
our universe, vacuum solutions to the classical Einstein equation generically
exhibit white holes and structures with noneuclidean topology, and there is a generically
large space of time-nonorientable solutions.  There are also vacuum solutions 
(and positive energy solutions) with closed timelike curves, though how generic these solutions 
are remains an open question. The absence of white holes is a central mystery, tied to the
thermodynamic arrow of time.  The absence of macroscopic topological structures and of 
macroscopic closed timelike curves (CTCs) is a similarly central
feature of our experience.  Where have they gone?  Why is the topology
and the causal structure of spacetime ``ordinary'' on a macroscopic
scale, when what we call ordinary is extraordinary in the space of
solutions?

Two entirely different answers are consistent with our knowledge.  The
first is simply that the classical theory has a much broader set of
solutions than the correct theory of quantum gravity.  It is not
implausible that causal structure enters in a fundamental way in
quantum gravity and that classical spacetimes with closed timelike
curves approximate no quantum states of the
spacetime geometry.  Less likely to us is the analogous
explanation for trivial spacetime topology, that the correct theory of quantum
gravity allows only euclidean topology and forbids
topology change.

A second possible answer is provided by topological censorship and
chronology protection:  One supposes that quantum gravity allows
microscopic (near Planck-size or near string-size) structures that 
have nontrivial topology and/or violate causality; and one shows that classical general
relativity and the character of macroscopic  matter (described by
classical or semiclassical fields) forbid exotic structures that are
macroscopic in their spatial and temporal size. 

This brief review emphasizes a few related areas of recent work and is not 
intended to be comprehensive in any sense.  
A monograph by Visser~\cite{visserbook} reviews much of the work prior to 1995 and supplies a 
comprehensive bibliography; informal reviews are given by Thorne~\cite{thornebook} and   
Gott~\cite{gottbook} (whose bibliography has brief, useful descriptions of each 
reference).  For a recent review of work on energy inequalities, 
see,for example, Roman~\cite{Romrev}, and an earlier popular article by 
Ford and Roman~\cite{fordroman}.  Earlier reviews by one of the present authors are
Refs.~\cite{jftexas,jfcanada}. For a more detailed (and more technical) review of 
the Cauchy problem on spacetimes with CTCs and on Lorentzian universes-from-nothing,
see Ref.~\cite{jf03}. Becasue most of chronology protection is concerned with isolated 
regions of CTCs, and because of space limitations, the extensive work on
Gott spacetimes is not covered here and no attempt is made to provide a 
comprehensive bibliography. Interested readers should balance Gott's 
view \cite{gottbook} with articles constraining CTC formation in 
3-dimensional spacetimes, beginning for example, with work by Deser, Jackiw 
and 't~Hooft \cite{djt92},  Cutler \cite{cutler92}, Carroll, Fahri and Guth \cite{cfg94} 
(and references therein), and Tiglio \cite{tiglio98}.

\section{Topological censorship}

\subsection{Expectation of nontrivial topology}
 
Beyond the wormholes that science fiction has made a part 
of popular culture 
lie an infinite variety of topological structures in three dimensions, 
a countably infinite set of prime 3-manifolds.  Witt~\cite{witt86} 
showed that {\sl all} 3-manifolds (prime and composite) occur as spacelike hypersurfaces 
of vacuum spacetimes -- solutions to the vacuum Einstein equations.  This is not enough to 
show that they arise as isolated structures: A flat universe with the topology of 
a 3-torus satisfies the field equations, but its existence does not imply that 
one can find a vacuum geometry for which the universe looks everywhere like a 3-sphere 
except in an isolated region.  Isolated systems are ordinarily modeled as asymptotically 
flat spacetimes; and a recent result by Isenberg 
et al.~\cite{isenberg03} shows that all 3-manifolds do in fact occur as isolated structures, 
as spacelike hypersurfaces of asymptotically flat, vacuum spacetimes. 

Whether topological structures can arise from a spacetime that is initially topologically 
trivial is a more difficult question.  Given any two 3-manifolds, 
$S_1$ and $S_2$, one can always find a spacetime (a 4-manifold with 
Lorentzian metric) that joins them and for which they are each spacelike: There is a 
spacetime whose spacelike boundary is the disjoint union of $S_1$ and 
$S_2$~\cite{reinhart63,sorkin86}.
(A strengthened version of this and a review of earlier work on classical topology change is 
given by Borde~\cite{borde94}). But one pays a price for topology change.  A theorem due to 
Geroch~\cite{geroch} shows that a spacetime whose boundary comprises two disjoint spacelike 3-manifolds always has 
closed timelike curves. (See \cite{borde94} for a version appropriate to asymptotically flat spacetimes.)  

Topology-changing spacetimes must also have negative energy.  Much of the recent work involving 
restrictions on spacetime topology and on causal structure relies on the null energy 
condition (NEC): An energy-momentum tensor 
$T^{\alpha\beta}$ satisfies the {\sl null energy condition} if, for any null vector $k^\alpha$, 
$T_{\alpha\beta}k^\alpha k^\beta \geq 0$.  
In particular, Tipler shows that topology-changing spacetimes violate the null energy
condition~\cite{tipler}.

Theorems involving the null energy condition use it to infer increasing convergence of 
light rays. Their proofs require only that the average value of 
$T_{\alpha\beta}k^\alpha k^\beta $ along null geodesics $\gamma$ be nonnegative. This weaker requirement, the {\sl averaged null energy condition} (ANEC), has the form  
\be
 \int_\gamma d\lambda\ T_{\alpha\beta}k^\alpha k^\beta >0,
\label{anec}\ee
with $\lambda$ an affine parameter along the null geodesic and $k^\alpha$ the corresponding 
null tangent vector.

Because classical matter satisfies the null energy condition, it seems unlikely 
that one can create topological structures with a time evolution that is nearly classical.  In path-integral approaches to quantum gravity, however, the amplitude 
for a transition from one 3-geometry to another is ordinarily written as a sum over all interpolating 
4-geometries.  Whether one works in a Euclidean or Lorentzian framework, topology change is thus permitted, if only at the Planck scale.  One can then ask whether small-scale topological structures can grow to macroscopic size and persist for macroscopically long times.

The meaning of topological censorship is that they cannot: Isolated topological structures 
with positive energy collapse, and they do so quickly enough that light cannot traverse them. 

\subsection{Gannon's Singularity Theorem}
The first result of this kind was due to Dennis Gannon~\cite{gannon}, 
a similar result obtained independently by Lee~\cite{lee}.
\newtheorem{thm}{Theorem}
\begin{thm}  Let $M,g$ be an asymptotically flat
spacetime, obeying the null energy condition,
\be 
T_{\alpha\beta}k^\alpha k^\beta\geq 0.
\ee
If $M,g$ has a nonsimply connected Cauchy surface, then it is
geodesically incomplete.
\end{thm}
Given the recently proven Poincar\'e conjecture~\cite{perelman,morgantian,caozhu},
Gannon's theorem implies that if the topology of the Cauchy surface $S$ is not trivial 
(i.e., if adding a point at infinty to $S$ yields any closed 3-manifold other 
than $S^3$), then the spacetime is geodesically incomplete. 

Requiring that the system be isolated is essential to the theorem.  Topological structures large enough 
to expand with the Hubble expansion are not ruled out, and the topology of the 
universe is unrestricted.  Each of the countably many hyperbolic, spherical, and flat 
3-manifolds is consistent with the homogeneity and isotropy of the observed universe. In fact, only with the high angular resolution of the recent microwave anisotropy probes has it been possible to show that the apparent size of the visible universe 
is not an illusion arising from light traversing several times a space whose size is a fraction 
the Hubble length~\cite{cornish04} (for a review, see Levin~\cite{levin02}). 
\footnote{The geometries of the early and present universe are apparently deSitter, 
and work by Witt and Morrow~\cite{wittmorrow} shows that 
every 3-manifold occurs as a slice of a deSitter 4-geometry -- though not in general  
a slice consistent with our universe.} 

The geodesic incompleteness proved by the theorem is the way most 
singularity theorems are stated.  It is expected in this context to imply 
gravitational collapse, leading (within the classical theory) to unbounded curvature within a 
black hole: Geodesics would then be incomplete because they reach 
the singularity (that is, scalars locally constructed from the curvature grow without bound) 
within finite affine parameter length.  The topological censorship theorem 
reinforces this expectation and confirms part of it. 
 
\subsection{Topological Censorship}

The conjecture can be stated in two equivalent 
forms~\cite{fsw,galloway95}.
Recall that to define a black hole -- a region from which light cannot escape -- 
one first makes precise the notion of a light ray reaching infinity by attaching 
to spacetime a boundary, future null infinity (${\cal I}^+$)~\cite{hawkingellis}.  
Light that is not trapped is then light that reaches future null infinity, and a black 
hole is the region from which no future-directed null geodesic reaches infinity.
Past directed null geodesics that are not trapped similarly 
reach past null infinity. (Unless the spacetime has a 
white hole, no past-directed light ray will be trapped).  An observer that 
can communicate with the outside world is one whose past and future directed null rays 
reach null infinity.  She is then outside all black (and white) holes, 
in the {\em domain of outer communication}.  We also need the term {\sl causal curve}, 
a curve whose tangent is everywhere timelike or null.

\begin{thm}{Topological Censorship Version A} (Friedman, Schleich,
Witt).  Let $M,g$ be an asymptotically flat, globally hyperbolic
spacetime satisfying the averaged null energy condition.  Then every causal curve
from past null infinity to future null infinity can be deformed to a curve
near infinity. (More precisely, each causal curve can be deformed with its endpoints fixed at 
${\cal I}$ to a curve that lies in a simply connected neighborhood of ${\cal I}$.)
\end{thm}
\noindent
That is, no causal curve can thread the topology. 
The theorem implies that no 
observer who remains outside all black holes (and who did not emerge from a 
white hole) can send a signal that will probe the spacetime topology.  

Topological censorship can be regarded as a statement about the topology of 
the domain of outer communications: The region outside all black (and white) holes
is topologically trivial.

\begin{thm}{Topological Censorship Version B} (Galloway).  Let
$M,g$ be an asymptotically flat spacetime obeying the averaged null energy 
condition, and suppose the domain of outer communication is globally 
hyperbolic.  Then the domain of outer communications is simply connected. 
\label{topcenb}\end{thm}
Again, given the Poincar\'e conjecture, simply connected is equivalent to 
topologically trivial.  Here this follows from the fact that a globally 
hyperbolic spacetime has topology $S\times {\bf R}$. Topological censorship  
implies that $S$ is simply connected, and the Poincar\'e conjecture then implies 
that $S$ has trivial topology, whence $ S\times {\bf R}$ has trivial topology.

The proof of Version A that is simplest to outline relies on an
argument used by Penrose, Sorkin and Woolgar\cite{psw93} in their proof of a
positive mass theorem.  Suppose a causal curve $\gamma$ joining ${\cal
I}^-$ to ${\cal I}^+$ is not homotopic to an asymptotic curve.  Denote
by $\lambda^\pm$ the generators of ${\cal I}^\pm$ that contain the
endpoints of $\gamma$.
Partially order all curves from $\lambda^-$ to $\lambda^+$, writing
$$\gamma_2 \geq \gamma_1, \quad {\hbox{if $\gamma_2$ is faster than
$\gamma_1$}}.$$
That is, $\gamma_2 \geq \gamma_1$ if $\gamma_2$ leaves $\lambda^-$ later
than $\gamma$ (or at the same time) and if $\gamma_2$ reaches $\lambda^+$
earlier (or at the same time).  Then a fastest curve $\gamma_\infty$ in
the homotopy class of $\gamma$ is a null geodesic without conjugate
points.  

But the Raychandhuri equation together with the null energy condition
implies that null geodesics have conjugate points in finite affine
parameter length, a contradiction.\\

\subsection{Implications for black holes}

Chru\'sciel and Wald showed that, in form B (Theorem \ref{topcenb}), topological censorship 
implies that stationary black holes have spherical topology~\cite{cw}. The result 
itself is due initially to Hawking~\cite{hawking}, with a proof that assumes 
analyticity. The proof based on topological censorship strengthens the theorem: 
It does not require analyticity (or smoothness) and it uses the weak energy condition.  
The proof given in Ref.~\cite{cw} was written before Galloway's version
B appeared, and by reading Galloway first, one can avoid some of the
detailed arguments in~\cite{cw}. The proof uses 
the fact that the domain of outer 
communication is simply connected to show that its boundary is homeomorphic to the
boundary of a compact three-manifold whose interior is 
simply connected.  This is enough to prove the theorem, because a standard
result for 3-manifolds asserts that such a boundary is a disjoint
union of spheres.

The result has been amplified in several directions~\cite{jv,bg94,bg95,gsww99,gsww01,gw01}.
In particular, one can dispense with stationarity to show spherical topology for 
slices of the horizon by Cauchy surfaces that lie to the future of a slice of 
past null infinity~\cite{bg94,bg95} or whose topology is unchanging~\cite{jv}.  
  
At first sight, it would seem that black holes must always have spherical 
topology, because the event horizon is the boundary of the {\sl simply connected} domain 
of outer communication.  This would match the intuitive picture of their formation and 
the familiar pair-of-pants coalescence of two black holes. A typical slicing 
of the horizon begins with two disjoint points that expand to two disjoint spheres; 
coalescence begins with the intersection of the two spheres in a single point; and 
the final slices are single spheres.  

In fact, however, as Hughes {\sl et al.} first found 
numerically~\cite{hkwwst}, one can find examples of collapse in which the intersection of 
a spacelike hypersurface with the horizon is toroidal. Examples of horizons with slices 
with higher genus are not difficult to construct, and work by Siino \cite{siino98} 
shows that slicings with arbitrary genus can be constructed when the horizon 
has caustics.  The reason is that the past endpoints of the horizon's 
null generators form an acausal set; and it is the intersection of wiggly spacelike 
slices with this acausal set that can have arbitrarily high genus.   
   
     It nevertheless appears that there is always an alternative slicing of the horizon 
in which black holes are spherical. That is, Siino proves the following result:  
Let $M,g$ be a strongly causal spacetime with an event horizon that is a smooth 
$S^2$ to the future of some spacelike hypersurface. Then the domain of outer 
communications of $M,g$ can be foliated by spacelike slices, each of which intersect 
the future horizon in a union of disjoint points and of spheres that are either 
disjoint or that intersect in points.  The assumption of a final smooth $S^2$ 
is related to the unproved cosmic censorship conjecture, and one would like to replace 
that assumption by a positive energy condition.   

Underlying the conjecture is the way a toroidal black hole adheres to 
topological censorship: The torus closes before light can traverse it, and in the 
examples we know of, that constraint appears to allow a foliation in which black holes 
are spherical.  A simple example, suggested 
(in one lower dimension) by Greg Galloway, is similar to one given by Shapiro, Teukolsky and Winicour~\cite{stw95} (see also \cite{hw99}). It is a null surface in Minkowski space 
that allows an initially toroidal slicing. 
(One can construct an artificial
spacetime for which this null surface is the event horizon,
but the construction is unrelated to the nature of the 
null surface and its slicings.) The surface is generated by the future-directed outward 
null rays from the spacelike disk $t=0,\ z=0,\ x^2+y^2\leq 1$, with $t,x,y,z$ standard Minkowski 
coordinates. The surface is rotationally symmetric about the $z$-axis. A spacelike surface that 
cuts through the null rays and then the disk, before dipping below the disk gives a toroidal slice.
And spacelike surfaces that lie below the edge of the disk and go upward at the center of the disk 
give spherical slices.  

\subsection{Lorentzian universes from nothing}

Universes whose spatial slices are compact 3-manifolds are finite in space, but have no boundary.
One can similarly constuct 4-manifolds with no past boundary that are finite in time -- {\sl 
universes from nothing}.   In a Euclidean framework for quantum gravity, manifolds of this 
kind arise in the Hartle-Hawking wavefunction of the universe.  An example with a Lorentzian 
metric and with CTCs is given by Gott and Li~\cite{gottli}; but CTCs are not an essential feature 
of Lorentzian universes from nothing.  For a large class of these spacetimes, 
one can always choose metrics without CTCs; time nonorientability is then their only causal 
pathology.  They are the only examples of topology change in which one has a
smooth, nondegenerate Lorentzian metric without closed timelike curves.
The double-covering space of these spacetimes is globally hyperbolic, and that fact 
implies the existence of generalized Cauchy surfaces.  

A simple example of such a spacetime is a M\"obius strip, whose median circle is 
taken to be a spacelike hypersurface.  Initial data on this median circle consists of 
a specification of a field and its gradient, and the initial value problem is well defined 
when the metric has no CTCs.  The M\"obius strip can be constructed from a cylinder  
with coordinates $t,\phi$, by the antipodal identification 
$t\rightarrow -t, \phi\rightarrow \phi+\pi$.  In four dimensions, antipodally-identified
de~Sitter space 
is an example that is locally indistinguishable from ordinary de~Sitter.  An initial value 
surface in this case is the antipodally identified 3-sphere at $t=0$, and some initial value surface 
passes through each point of the spacetime.      

More generally, one can construct a Lorentzian universe with no past boundary from 
any compact 3-manifold $S$ that admits a diffeomorphism $I: S\rightarrow S$ 
by an analogous antipodal identification of $S\times{\bf R}$.  If one begins with 
antipodally symmetric initial data for the Einstein equations on $S$, then the 
resulting solution to the field equations on $S\times {\bf R}$ induces a metric on 
the identified space that is also a solution~\cite{fh,f98}. 
 
Because classical physics on Lorentzian universes from nothing has no pathology, it 
is natural to ask if there is any reason why
these 
spacetimes are forbidden 
if one incorporates quantum physics.  
Kay \cite{Kay2} proposes to impose the standard 
canonical commutation relations in a neighborhood of any point with
respect to one time orientation (the ``F-locality condition").  This
condition rules out time non-orientable spacetime in an obvious manner.
Gibbons~\cite{gibbons1} also rules out conventional quantum field theory
based on a complex Hilbert space in some time non-orientable spacetimes,
including antipodally identified de~Sitter spacetime.  

Refs.~\cite{fh,f98} investigate whether quantum field
theory in Lorentzian universes from nothing could be defined in 
globally hyperbolic neighborhoods with suitable overlap conditions on the intersections.
It turns out that there are no difficulties in defining an algebra of
fields in this manner.  It is also possible to construct states
satisfying the positivity conditions in each of the neighborhoods if
the union of two neighborhoods is always time orientable. 
This restriction on the neighborhoods, however, is rather artificial because there
will always be sets of points in the spacetime for which no correlation
function is defined.   If one then 
allows the union of two neighborhoods on which field theory is defined
to be time non-orientable, one can show that there are no physically reasonable
states satisfying the postivity condition in each neighborhood.
One may be able to construct a consistent field theory in these
spacetimes using path-integral quantization, but this possibility
has not been explored.

\section{Chronology protection}
\subsection{Overcoming the grandfather paradox: 
     Existence of solutions for generic data in a class
     of spacetimes with CTCs}

Closed timelike curves were traditionally regarded as unphysical because 
of the {\sl grandfather paradox}, the fact that a time 
machine would allow 
one to go back in time and do away with her grandfather.  More precisely, 
it was thought that most initial data on a spacelike surface would fail 
to have a consistent evolution: Locally evolving the data would not lead 
to a global solution, because the local solution would be inconsistent with 
the data, once the evolution returned to the initial surface.

A simple example of a spacetime for which generic initial data has no solution is 
the cylinder obtained from the slab of Minkowksi space between 
$t=-1$ and $t=1$ by identifying the points $-1,x,y,z$ and $1,x,y,z$. Data on 
the surface $t=0$ for, say, a massless scalar field $\psi$ can be locally 
evolved around the cylinder (to $t=1$ and then from the identified surface 
$t=-1$ back to $t=0$). The locally evolved solution, however, returns to the 
surface with a value of the field that disagrees with its initial value 
unless the initial data is chosen to give a solution periodic with period $T=2$.

One of the surprises that spurred interest in spacetimes with CTCs is the existence 
of a different class of spacetimes for which solutions exist for free fields with 
arbitrary initial data, for interacting classical particles (billiard balls), 
and perhaps for interacting fields as well.  
Whether physics is consistent with closed timelike curves, however, does not 
rely on consistency of classical physics.  A path-integral assigns probability amplitudes to histories, whether or not 
there are classical histories.  And in the billiard-ball case, where classical solutions 
exist but are not unique, a path-integral dominated by classical solutions would simply 
assign large amplitudes to the alternative classical histories. 

We review the billiard-ball example and interacting field arguments briefly in 
the next section, after we have introduced another classical obstacle to closed 
timelike curves (an instability of the Cauchy horizon) that these examples overcome.
In this section we consider free fields.

\begin{figure}
\includegraphics[height=3in]{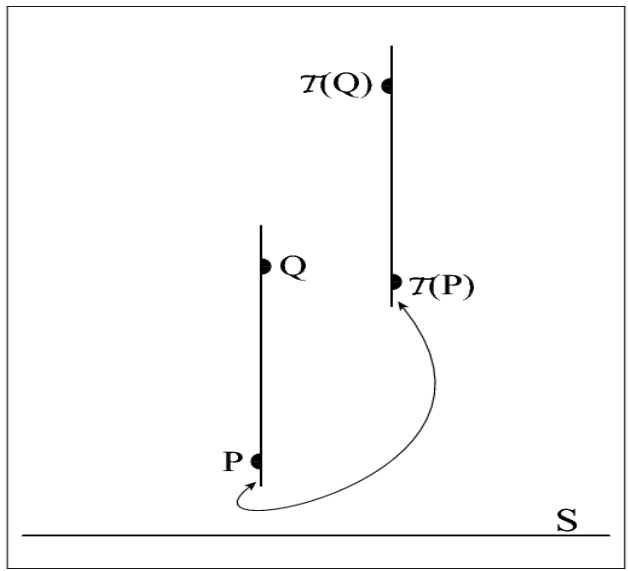}
\caption{\label{slits} A simple spacetime with CTCs and a generalized Cauchy 
surface $S$ is shown in this figure.
(The precise definition of a generalized Cauchy surface is given below.)
Two parallel segments of equal 
length are removed from Minkowski space, two disjoint edges are 
joined to the left and right sides of each slit, and edge points 
related by the timelike translation ${\cal T}$ are then 
identified. }
\end{figure}

Two-dimensional spacetimes built from Minkowski space 
that avoid the grandfather 
paradox are easily constructed.  An example, similar to 
spaces discussed by Geroch and Horowitz~\cite{gh} and by
Politzer~\cite{politzer1}, is constructed by removing two parallel timelike slits 
from Minkowski space and gluing the edges of the slits. Corresponding points on the two inner 
edges are identified 
by the translation $\cal T$ shown in Fig.~\ref{slits}; and corresponding points on the two outer edges are similarly identified.

CTCs join identified points of the inner edges, from $Q$ on the left to 
${\cal T}(Q)$ on the right.  The hypersurface $S$, lying to the past 
of the CTCs, is an obvious candidate for a generalized
Cauchy surface of $M,g$.\\

{\em Definition}. A {\em generalized Cauchy surface} $S$ is an 
achronal hypersurface of $M$ for which the initial value problem for 
the scalar wave equation is well-defined: 
Any smooth data in $L_2(S)$ with finite energy, for a scalar field $\Phi$,
has a unique solution $\Phi$ on $M$.  \\

In fact, it is easy to see that initial data
in $L_2(S)$ leads to a solution in $L_2(M)$.  In the past of the CTCs 
(in the past of the Cauchy horizon),
solutions to the massless wave equation can be written as the sum
$f(t-x) + g(t+x)$ of right-moving and left-moving solutions.  To
obtain a solution in the spacetime $M$, one simply propagates left
moving data that encounters the slit in the obvious way.  For example,
if a left-moving wave enters the left slit at $Q$, it emerges unaltered
from the right slit at ${\cal T}(Q)$.  The solution is unique. But it is
discontinuous along future-directed null rays that extend from the
endpoints of the slits, because the result of the wave propagation is
to piece together solutions from disjoint parts of the initial data
surface (see Friedman and Morris~\cite{fm97}).  
Analysis of the initial value problem for a related spacetime with
spacelike slits is given by Goldwirth {\sl et al.}~\cite{gppt}.

Although, in our 2-dimensional example, the solution has discontinuities,
in  four dimensions one can construct spacetimes for which smooth, 
unique solutions to the scalar wave equation 
exist for all data on a generalized Cauchy surface. 
%
%
%
The examples for which these statements are proved are time-independent 
spacetimes~\cite{fm97,bachelot}, in which CTCs are always present.  
The spacetimes are asymptotically flat, and one can
define future and past null infinity. In Minkowski space past null infinity is a 
generalized Cauchy surface for massless wave equations, and the theorems show 
that it is also generalized Cauchy surface for a class of spacetimes with CTCs.   
 
\begin{figure}
\includegraphics[height=2in]{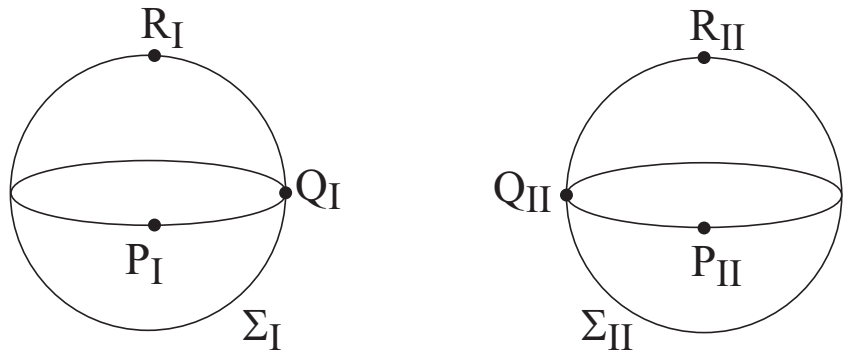}
\caption{\label{tunnel} A wormhole is constructed by removing two balls and 
identifying their spherical boundaries $\Sigma_I$ and $\Sigma_{II}$ after 
a reflection: Points labeled by the same letter, with subscripts I and II, 
are identified. }
\end{figure}

\begin{figure}
\includegraphics[height=3in]{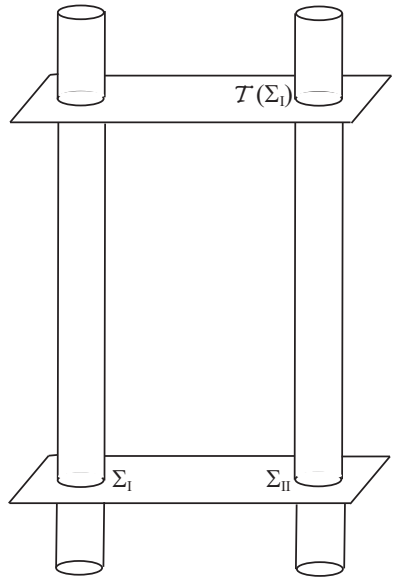}
\caption{\label{cylinders} The sphere labeled $\Sigma_1$ is identified with 
the timelike separated sphere at ${\cal T}(\Sigma_I)$}
\end{figure}

The spacetime considered in Ref.~\cite{fm91,fm97} has a wormhole that joins points at 
one time to points at an earlier time.  Recall that a wormhole is 
constructed 
from ${\bf R}^3$ by removing two balls and identifying their spherical boundaries,
$\Sigma_I$ and $\Sigma_{II}$, as shown in Fig.~\ref{tunnel}. 
The history of each sphere is a cylinder in spacetime, and one constructs 
a spacetime with CTCs by removing two solid cylinders from Minkowski 
space and identifying their boundaries $C_I$ and $C_{II}$ after a time 
translation, so that the sphere at time $t+T$ is identified with a sphere 
at an earlier time $t$, as shown in Fig.~\ref{cylinders}. Thus a particle 
entering the wormhole mouth at ${\cal T}(\Sigma_I)$ emerges from 
the mouth $\Sigma_{I}$ at an earlier time.\\
 The existence proof relies on 
a spectral decomposition of the field. Technical difficulties arise 
from the fact that the boundary conditions are frequency dependent, and 
solutions with different frequencies are not orthogonal. As a result, one 
cannot use the spectral theorem, and a separate proof of convergence of the 
harmonic decomposition is required.  Recent work by Bugdayci \cite{bugdayci} constructs 
a solution for the case of a metric that is flat everywhere outside the 
identified cylinders, as a multiple scattering series. 
 
More recently, Bachelot~\cite{bachelot} proved a similar existence
theorem and a strong uniqueness theorem for another family of
stationary, four-dimensional spacetimes that are flat outside a
spatially compact region.  These spacetimes have Euclidean topology and
their dischronal regions have topology (solid torus) $\times {\bf R}$.  The
metric is axisymmetric, with one free function $a$ that describes the
tipping of the light cones in the direction of the rotational Killing
vector $\partial_\phi$.

\subsection{Classical chronology protection}

Parallel slits do not exhibit a property of almost all other 
two-dimensional spacetimes whose CTCs lie to one side of a Cauchy horizon: 
a classical instability of the Cauchy horizon.  Once the slits 
are not parallel (once they are, in effect, walls in relative motion),
the instability arises.  The paradigm spacetime 
for this instability is 
{\em Misner space}~\cite{misner,hawkingellis,thorne94}.
 
\begin{figure}
\includegraphics[height=3in]{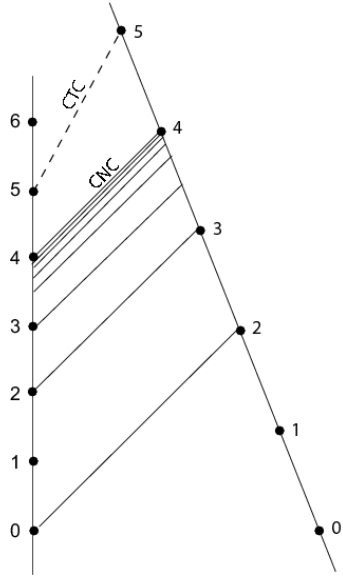}
\caption{\label{fig:misner} Misner space is the region between the two 
identified timelike lines.  A solid null ray loops around the space, 
approaching a closed null geodesic marked CNC. 
Closed timelike curves (one labeled CTC) pass through each 
point above the closed null geodesic. }
\end{figure}

Misner space can be obtained from a 1-dimensional 
room whose walls are moving toward each other at relative speed $v$, 
by identifying left and right walls at the
same proper time read by clocks  on each wall. 
The resulting space is the piece of Minkowski space between 
two timelike lines (the walls), with the lines identified by the boost that 
maps one line to the other.  In the diagram below,  Fig.~\ref{fig:misner}, numbers 
label readings of a single clock, shown at identified points of the left and 
right walls.  A light ray beginning at the left wall at $t=0$ is boosted each 
time it traverses the space, in the same way that light is boosted when reflected 
by a moving mirror.  As it loops around the space, the light ray approaches 
a closed null geodesic through the clock at $t=4$.  

Identified points are spacelike separated for $t<4$. For $t=4$ the 
identified clocks are separated by a null geodesic that marks the 
boundary of the globally hyperbolic spacetime to its past and the 
{\em dischronal} region above it.  Through each point of the 
dischronal region passes a closed timelike curve; an example is  
the dashed line joining the clock images at $t=5$.

The divergence of solutions to the wave equation is clear in the geometrical optics limit.
A family of light rays that loop about the space are boosted at each loop, their 
frequencies increased by the blueshift factor $[(1+v)/(1-v)]^{1/2}$. Their 
energy density, measured in the frame of any inertial observer, diverges as 
they approach the Cauchy horizon. Each looping light ray is an incomplete null 
geodesic: It reaches the horizon in finite affine parameter length, because each boost decreases
the affine parameter by the factor $ [(1+v)/(1-v)]^{1/2}$. 
In discussing Gannon's singularity theorem for nontrivial topology, we noted that, 
in the context of gravitational collapse, geodesic incompleteness is thought generically 
to imply a curvature singularity. Here, however, the spacetime is smooth, and the incomplete geodesic is unpleasant only because it leads to an instability of the Cauchy horizon. 

This behavior is not unique to Misner space or to two dimensions: A theorem 
due to Tipler~\cite{tipler} shows that geodesic incompleteness is generic in
spacetimes like Misner space in which CTCs are ``created'' --
spacetimes whose dischronal region lies to the future of a spacelike
hypersurface.   The nature of the geodesic incompleteness is clarified 
by Hawking\cite{hawking}, at least in the case when the null generators of 
the Cauchy horizon have the character of a {\sl fountain}, all springing from 
a single closed null geodesic (past-directed generators appoach the closed 
null geodesic).  The boosted light rays that give rise to the Misner space instability will 
characterize these more general Cauchy horizons in four dimensional spacetimes.
In four dimensions, however, the instability competes with the spreading of 
the waves.  In the geometrical optics limit, the boosting will not lead to 
a divergent energy density if the area of a beam increases as the beam loops 
by a factor greater than the increase in energy density due to the boost 
in frequency.  That is, if the ratio $A_{n+1}/A_n$ of beam areas at successive loops 
is greater than $(\omega_{n+1}/\omega_n)^2$, the boosted frequency will not 
lead to a divergence of the energy at the Cauchy horizon. 

\begin{figure}
\includegraphics[height=3in]{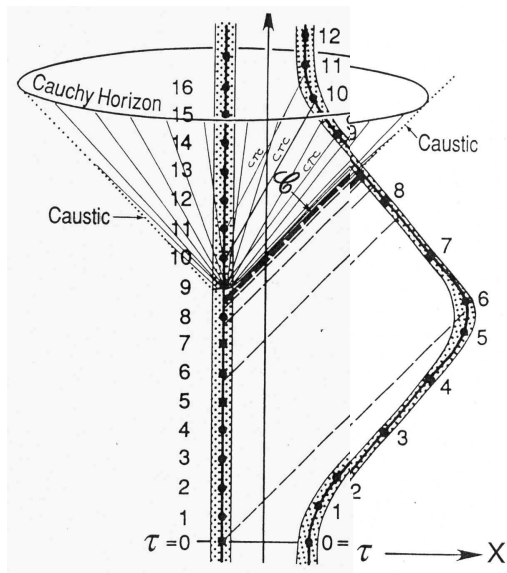}
\caption{The wormhole spacetime of Morris et al. is shown in this figure. The
two mouths at the same proper time are identified.}
\label{wormhole}
\end{figure}

Hawking's analysis follows an example given by Morris, Thorne and Yurtsever~\cite{mty,consort} 
of a wormhole whose mouths move toward each other in a way that initially mimics the Misner-space 
walls, as shown in Fig.~\ref{wormhole}. 
  
For the case of a spacetime flat outside the 
wormhole mouths (the identified spheres), each time a beam of light traverses the 
wormhole, its frequency is boosted and its area increases by the approximate factor 
$(d/R)^2$, with $d$ the distance between wormholes, $R$ the radius of the throat. 
In this example, the horizon generators do have the fountain behavior described above.
The fountains assumed by Hawking, however, are probably not generic for the kind of 
Cauchy horizons he considers (compactly generated, noncompact).  
(A Cauchy horizon is {\it compactly generated} if all its past-directed 
null generators enter and remain in a compact region.)
Chru\'sciel and Isenberg~\cite{ci} give examples of compactly generated 
noncompact Cauchy horizons whose generators do not have fountainlike behavior (for 
which no closed null geodesic is an attractor); they show that fountains are 
not generic for compact Cauchy horizons and argue that they are not generic for 
the compactly generated, noncompact case.

Because the evolution of fields on the wormhole spacetime does not lead to an instability 
of the Cauchy horizon, it appears that asymptotically flat spacelike hypersurfaces 
to the past of the Cauchy horizon (e.g.,  a $t=$ constant surface of Minkowski space 
through $\tau =0$ in Fig.~\ref{wormhole}) are generalized Cauchy surfaces. One can 
formally construct a solution as a multiple scattering series, and, for sufficiently 
small ratio $R/d$, we expect the series to converge. 

For interacting particles, modeled as billiard balls, 
Echeverria {\sl et al.}~\cite{echev} 
looked at the simpler time-independent wormhole spacetime of Fig.~\ref{cylinders}, in which  
billiard balls entering the right mouth of the wormhole at
time $t$ exit from the left mouth at time $t-\tau$.  Consider a ball 
that is aimed at the right wormhole along a timelike line that 
intersects itself. Initial data for the ball has a local solution that,
when extended has the exiting billiard ball aimed to strike its earlier 
self, apparently preventing a solution.  In fact, however,
there appear always to be {\em glancing blow} solutions for arbitrary
initial position and velocity of the incoming ball.  In these
solutions, the incoming ball is hit by its earlier self in just the
right way that it enters the wormhole and emerges aimed to strike
itself that glancing blow. \footnote{By adding an additional degree of freedom, 
however, one can apparently obtain systems that do not 
admit classical solutions~\cite{jftexas,everett}.}

With both the grandfather paradox and the classical Cauchy horizon instability 
overcome, what prevents the classical formation of spacetimes with CTCs?  In the 
wormhole example, the null energy condition must be violated, if one is to keep the 
wormholes open long enough for light to traverse them, and hence long enough for 
closed timelike curves to traverse them.  Hawking shows that this is generically 
true~\cite{hawking}.  \\ 


{\sl Classical chronology protection}.  Let $M,g$ be a spacetime with
a compactly generated Cauchy horizon. Assume that the Cauchy horizon 
is the boundary of the domain of dependence of a noncompact partial 
Cauchy surface $S$. Then the null energy condition is violated.\\

The Cauchy horizon is expected to be compactly generated
in a nonsingular,  asymptotically flat spacetime 
for which the Cauchy horizon bounds the domain of dependence of an 
asymptotically flat spacelike hypersurface $S$.
%
%
The argument is then:  \\
(i) Generators of a future Cauchy horizon are null geodesic
segments with no past endpoints.\\
(ii) Because they enter a compact region, some of the generators 
must converge to the past.  \\
(iii) Positive convergence for a past-directed generator, together with 
the  Raychaudhuri equation, implies that the null energy condition must 
be violated on the future horizon.  Otherwise the generators would 
have past endpoints.\\

If one allows the spacetime to be singular to the future of $S$, as in  
recent examples by Ori~\cite{ori}, it is not difficult 
to show that one can find chronology-violating  
asymptotically flat spacetimes satisfying the null energy condition. In this 
case the Cauchy horizon is singular and thus not compact (past directed 
null geodesics run off the manifold in finite affine parameter length).
In Ori's most recent example, a spacetime whose only matter is a compact 
region of dust has a Cauchy horizon with closed null geodesics that is 
nonsingular in a neighborhood of these geodesics.  

Because quantum fields fail to satisfy the null energy condition, a number of 
attempts have been made to circumvent the classical chronology protection 
theorem and topological censorship by finding a way to have negative 
energy regions that persist for long times; or to have solutions with CTCs 
or nontrivial topology that require only small amounts or small regions of 
negative energy.  In our opinion, however, the recent work on quantum energy 
inequalities, discussed in Sect.~\ref{sec:qei} below, has set increasingly stringent 
constraints on violations of 
the 
energy 
%
conditions. 
We think it likely 
that macroscopic violations of either topological censorship or classical 
chronology protection are inconsistent with the fundamental properties of 
semiclassical quantum fields.

\subsection{Quantum chronology protection}
Various pathologies have been found in quantum field theory 
in spacetimes with  
dischronal regions. A quantum instability in spacetimes with 
a Cauchy horizon may prevent the formation of CTCs on scales large compared to 
the Planck scale;  or, together with a loss of unitarity for interacting fields 
on spacetimes with CTCs, it may indicate 
that chronology violations do not 
occur in the fundamental theory.    
 
We begin with a brief heuristic summary of the quantum instability and then present a  
more technical description that includes more recent results.
  
\subsubsection{Quantum instability}

The classical instability of Misner space 
has a quantum counterpart that is present in cases where classical fields remain 
finite.  The quantum instability and the loss of unitarity are each related to 
propagation of fields around closed null or timelike curves. In computing the 
energy density of a quantum field in the vacuum, 
one must renormalize the field to produce a finite result, subtracting off a 
divergent zero-point energy of the vacuum. One can, for example, impose a 
short-distance cutoff, subtract a spacetime-independent term that would be the 
zero-point energy of a flat spacetime with a short-distance cutoff, and then 
take the continuum limit (as the cutoff goes to zero). The subtraction eliminates the 
divergence in the propagator from nearby points that are separated by a null 
geodesic.  When there are closed null geodesics, however, more than one null 
geodesic connects nearby points, and additional divergent contributions arise 
from vacuum fluctuations that propagate around these closed null curves. The 
result is that, as one approaches a Cauchy horizon, the finite, renormalized 
energy density can grow without bound.  

A divergence of the renormalized energy-momentum tensor at a Cauchy horizon, 
$\langle T^{\rm ren}_{\alpha\beta}\rangle$, was examined by various authors 
for free scalar field theory in several chronology violating 
spacetimes~\cite{hk,tanaka,kt,hawking,visser}.  
These results, together with the Hawking's energy-condition violation theorem of 
the previous section (classical chronology protection) led Hawking to 
propose his Chronology Protection Conjecture: {\em The laws of physics prevent
closed timelike curves from appearing}.  More precisely, Hawking
conjectured that the laws of physics prevent local {\em creation} of closed
timelike curves, as characterized by the existence of a
compactly generated Cauchy horizon.
In several examples,  however, the value of $\langle T^{\rm ren}_{\alpha\beta}(x)\rangle$
remains finite as the point $x$ 
approaches the Cauchy horizon~\cite{tanaka2,Krasnikov,sushkov,sushkov2}.  Subsequently Kay,
Radzikowski and Wald (KRW)~\cite{krw}
showed that the two-point function, from which the
energy-momentum tensor is obtained, 
is singular, in the sense which will be explained below,
for a free scalar field at some points on 
a compactly generated Cauchy horizon.  Thus, although 
$\langle T^{\rm ren}_{\alpha\beta}\rangle$ 
may be bounded as the point approaches the
Cauchy horizon, it is not well-defined {\em on the horizon}.

The KRW theorem uses the fact that for a Hadamard state the
two-point function 
$\langle \hat{\varphi}(x)\hat{\varphi}(x')\rangle$ is divergent if and
only if the two points $x$ and $x'$ can be connected by a null 
geodesic~\cite{radz1}.
There is now a wide consensus that a physically
reasonable state must be a Hadamard state (see, e.g. Ref.~\cite{Wald1}
for use of the Hadamard condition). 
This condition roughly states that the light-cone
singularity structure must be the same as for the vacuum state in
Minkowski spacetime.

In a globally hyperbolic spacetime every point $x$ has a 
convex normal neighborhood $N_x$~\cite{hawkingellis}
small enough that no null geodesic leaves and reenters it.
This implies for each point $x$ in a globally hyperbolic spacetime the following
property: For a small enough convex normal neighborhood $N_x$ and for
any two points $y, y'\in N_x$, the two-point function
$\langle \hat{\varphi}(y)\hat{\varphi}(y')\rangle$ is divergent if and
only if $y$ and $y'$ are connected by a null geodesic inside $N_x$.  
Because this property is necessary to define $T^{\rm ren}_{\alpha\beta}(x)$, 
we say that the two-point function is singular at $x$ if the property 
is not satisfied.

Now, for a compactly generated Cauchy horizon, every past-directed null 
geodesic generator $\lambda$ stays in a compact region. There must then be 
a point $x$ such that, given any
neighborhood $N_x$, the geodesic $\lambda$ passes through $N_x$
infinitely many times.  This implies that for any convex normal
neighborhood $N_x$ of $x$ there are points $y$ and $y'$ in $N_x$ such
that the two-point function $\langle
\hat{\varphi}(y)\hat{\varphi}(y')\rangle$ is divergent.  Hence we have
the following theorem due to KRW:
\begin{thm}
The two-point function of a free scalar field is singular on certain
points on a compactly generated Cauchy horizon.
\end{thm}
 For explicit verification of the KRW theorem in
examples with vanishing $\langle T_{\alpha\beta}^{\rm ren}\rangle$, 
see Ref.~\cite{ck1}.
Although KRW proved this theorem for a scalar field, it would be
straightforward to generalize their result to non-interacting 
fields of any spin.

Because of examples in which the stress tensor does not diverge, and because the 
theorem does not imply that the strong quantum instability of the earlier examples 
is generic, one can argue that the quantum singularity is too weak to 
enforce chronology protection.  Kim and Thorne~\cite{kt} had already entertained that 
argument in their strong-instability example, finally suggesting that the issue 
can be decided only within a theory of full quantum gravity.  
Visser~\cite{visserrely,visserproc} similarly argues that the above theorem
should be interpreted as the statement that quantum field theory in a
background spacetime is unreliable on the Cauchy horizon, and that
quantum gravity is needed to determine what really happens. 
The theorem does, however, indicate a drastic difference between 
the behavior of physical fields and spacetime on observed scales 
and their behavior on the smallest scales if there is to be 
chronology violation.

Kay's F-locality condition mentioned before
requires (among other things) that
quantum scalar fields in non-globally hyperbolic
spacetime commute for any two spacelike separated points 
in some neighborhood $N_x$ of any point $x$.  
The work of KRW showed that this condition cannot be satisfied in
spacetimes with a compactly generated Cauchy horizon.  Interestingly,
it can be satisfied on the timelike cylinder obtained as the 
quotient of Minkowski space by a 
time translation~\cite{Kay2,fhw}.
However, this was shown not to be the case 
for a massive scalar field in a two-dimensional spacelike
cylinder with a generic metric~\cite{fewsterold}.

\subsubsection{Loss of unitarity}

In the classical billiard-ball models mentioned above, it was found
that the Cauchy problem is not well defined because there is often more
than one solution for a given set of initial data~\cite{consort,echev}. 
In the corresponding quantum theory, however, it was found that the
solution of the wave equation is unique.  Although this observation
gave some hope that the quantum theory of interacting systems with CTCs might be
consistent, it turns out that unitarity is lost for interacting quantum 
fields in spacetimes with CTCs.  Ways to recover 
a consistent quantum theory have been 
suggested, but all have features that 
seem undesirable.

Unitarity of the scattering matrix $S_{fi} = \delta_{fi} - iT_{fi}$ in
quantum field theory can be expressed as
\be
2\,{\rm Im}\,T_{fi} = - \sum_{n}\overline{T_{nf}}T_{ni}\,,  \label{unirel}
\ee
As for free field theories, it was shown
in Ref.~\cite{fps} that,  if the Cauchy problem
is well defined for classical field equations, then these unitarity
relations are satisfied.  Thus, for example, the massless scalar field
theory in the spacetime with a chronology violating
wormhole studied by Friedman and Morris~\cite{fm97} is unitary.

The unitarity relations (\ref{unirel}) were studied for the
$\lambda\,\varphi^4$ theory in the above-mentioned wormhole spacetime by
Friedman, Papastamatiou and Simon~\cite{fps1} and 
in Gott spacetime by Boulware~\cite{boulware}.  They defined
the perturbation theory in a path-integral framework and found that
the Feynman propagator $i\Delta_F(x,y)$ has an extra imaginary part
$E(x,y)$:
\be
i\Delta_F(x,y) = \theta(x_0-y_0)D(x,y)+\theta(y_0-x_0)\overline{D}(x,y)
+ E(x,y)\,. \label{extra}
\ee
At first order in $\lambda$ the $T_{fi}$ in (\ref{unirel}) with the initial and
final states both being one-particle states corresponds to a tadpole
diagram.  The imaginary part of this diagram is essentially given by 
$E(x,x)$, which is non-zero.  This implies violation of
the relation (\ref{unirel}) because the
right-hand side starts at order $\lambda^2$.
In an unpublished work Klinkhammer and Thorne showed using the WKB
approximation that the quantum theory of the billiard
system with CTCs is non-unitary.  This non-unitarity arises due to the
fact that the classical system allows a multitude of solutions for a
given set of initial data and that the number of solutions depends on
the initial data.

Simple quantum mechanical models which mimic chronology violating
interacting field theory were studied by Politzer~\cite{politzer1}.  
He confirmed perturbative non-unitarity in a billiard model.  He also
studied some exactly solvable models exhibiting non-unitarity~\cite{politzer2}.
Fewster, Higuchi and Wells~\cite{fhw} studied a 
generalization of Politzer's model by solving the
differential equation satisfied by the Heisenberg operators.  They found
that the canonical (anti-)commutation relations are not preserved in
time, and, consequently, that the theory is non-unitary.   They also
compared their method and the path-integral quantization and
found that the two methods give different results.  This conclusion
seems to be related to the observation~\cite{FewWell} that the
Schr\"odinger and Heisenberg pictures do not agree in non-unitary
quantum mechanics.

Loss of unitarity poses difficulties for the conventional
Copenhagen interpretation of quantum mechanics.  Suppose, for example, 
one makes a measurement in a region spacelike separated from the CTCs.
One would expect that the result would be unaffected by the chronology
violation.  However, the measurement could be interpreted to have occurred
either before or after the CTCs, and the probability assignment would
depend on which interpretation was taken because of
non-unitarity~\cite{jacobson}.

There have been a few proposals for eliminating non-unitarity or for 
finding a probability interpretation that accepts loss of unitarity: 
Adopting only the unitary part of the evolution operator~\cite{anderson} and
making the non-unitary operator a restriction of a larger
norm-preserving operator~\cite{FewWell}; but the former would make
the evolution of states highly nonlinear, and the latter necessitates
the use of negative-norm states.  Friedman {\sl et al.}~\cite{fps1} and
Hartle~\cite{hartle} have advocated the
sum-over-histories approach to the interpretation of quantum mechanics.
This gives a prescription for computing probabilities, but probabilites in 
a globally hyperbolic {\em past} of any CTCs are affected by the existence 
of CTCs in the future.    
Hawking~\cite{hawk} has advocated the use of the superscattering
operator, i.e. the linear mapping from the initial to final density
matrices.  However, Cassidy~\cite{cassidy}
has found that the initial pure state evolves
nonlinearly into a mixed state.  Also, non-unitarity in the models studied in
Ref.~\cite{fhw} is such that evolution cannot even be described by a
superscattering matrix.  

Let us conclude this subsection by describing the work of Deutsch on
the grandfather paradox in
the context of quantum information theory~\cite{deutsch}.  Let the
initial state containing the grandfather of the killer be
$|\psi\rangle$ and let $\hat{\rho}$ be the density matrix resulting from
the evolution of
$|\psi\rangle\langle\psi|$ through the CTC region,
which may or may not contain the killer.  (Deutsch asserts that a
generic pure state would inevitably evolve to a mixed state, violating
unitarity, if there was a closed timelike curve.)  The state
$|\psi\rangle\langle\psi|$ in a Hilbert space ${\cal H}_1$
and its future self, $\hat{\rho}$, in a Hilbert space ${\cal H}_2$ form a
tensor product state $|\psi\rangle\langle \psi|\otimes \hat{\rho}$
in ${\cal H}_1 \otimes {\cal H}_2$ when they encounter one another.
The interaction of the two parts is described by a unitary matrix $U$ on
${\cal H}_1 \otimes {\cal H}_2$.  After the interaction, the Hilbert
space ${\cal H}_2$
continues to the future and ${\cal H}_1$ goes back to the past.  The
state going back to the past is obtained by tracing out the
resulting state over ${\cal
H}_2$ and must equal $\hat{\rho}$.  Thus, the consistency condition is
${\rm Tr}_2 \left[ U(|\psi\rangle\langle\psi|\otimes 
\hat{\rho})U^\dagger\right] = \hat{\rho}$.
Deutsch showed that there are 
solutions to this equation for any initial state $|\psi\rangle$ in
simple examples.  However, as we have seen, the evolution in chronology
violating spacetimes cannot be described in general by a linear
superscattering operator as envisaged in these models.

\section{Quantum energy inequalities}
\label{sec:qei}
\subsection{Introduction}

In quantum field theory, the weak energy condition
is violated by an energy-density 
operator whose vacuum expectation value is unbounded from below in Minkowski space, 
even for a free scalar field.  The energy condition is recovered in the classical 
limit, because the terms responsible for its violation oscillate 
%
in time 
and a classical field measures an average of the fluctuating quantum field.    
This averaging argument leads in two related ways to energy conditions satisfied by quantum fields: It suggests that the {\em averaged} null energy condition, (\ref{anec}), may still be valid; the ANEC is sufficient for establishing chronology protection and topological censorship (as well as some singularity theorems~\cite{Tipler2,Borde,Roman1}).  
And it suggests that a weighted average of the energy density
along a timelike curve may be bounded from below.  
Indeed, in a Minkowski-space context, Ford and Roman found such 
bounds~\cite{Ford2,fr1,fr2}, commonly called quantum (weak)
energy inequalities (QEIs). In the next subsection we briefly discuss some
work related to the ANEC in quantum field theory and then review
recent developments in QEIs. (See Refs.~\cite{Romrev,fewrev} for more
comprehensive reviews of the QEIs.)

First, however, we present a simple example of the violation of the weak energy 
condition by a free scalar field in Minkowksi space. The example relates 
the violation to the subtraction of an infinite constant from a formally positive 
energy-density operator $T_{00}$; and it exhibits the 
%
oscillation in time that leads to averaged inequalities.    

The classical energy density $T_{00}$ for a real 
massless scalar field $\varphi$ in four dimensions is
\be
T_{00} = \frac{1}{2}\left[ \dot{\varphi}^2  + (\nabla
\varphi)^2\right]\,.
\ee
To define the renormalized energy-density operator, one subtracts the infinite 
vacuum energy density, $\langle 0|\hat{T}_{00}|0\rangle$.  This renormalized 
operator is no longer positive-definite, although the energy, its spatial 
integral, is. One therefore expects states for which
the expectation value of the renormalized energy-density operator is
negative, and this is indeed the case~\cite{egj,Ford1}, as seen in 
the following simple example.
Let $a^\dagger({\bf p})$ be the creation operator for the scalar
particle with momentum ${\bf p}$ and consider the normalized superposition 
of the vacuum state and a two-particle state,
\be
|\psi\rangle = \cos\alpha |0\rangle + 
\frac{\sin\alpha}{\sqrt{2}}
\left[\int \frac{d^3{\bf p}}{(2\pi)^32p}
f({\bf p})a^\dagger({\bf p})\right]^2|0\rangle
\ee
with $p\equiv \|{\bf p}\|$ and
\be
\int \frac{d^3{\bf p}}{(2\pi)^32p}[f({\bf p})]^2 = 1\,,
\ee
where the function $f({\bf p})$ has been assumed to be
real for simplicity.
For example, if the function $f({\bf p})$ is chosen to be peaked about 
${\bf p}=\overline{\bf p}$ by letting
$f({\bf p}) = (12\pi^2 \overline{p}/\delta^3)^{1/2}$ if 
$\|{\bf p}-\overline{\bf p}\| < \delta$ 
and zero otherwise, then one finds
the expectation value of the renormalized energy-momentum tensor
at the origin for $\delta \ll \overline{p}$ as follows:
\be
\langle \psi|\hat{T}_{00}^{\rm ren}(t,0)|\psi\rangle
\approx \frac{\overline{p}\delta^3}{6\pi^2}\sin\alpha
\left( - \sqrt{2}\cos\alpha\,\cos 2\overline{p} t + 2\sin\alpha\right)\,.
\label{simpleeq}
\ee
This quantity is negative at $t=0$, say, if 
$\sin\alpha(\cos\alpha-\sqrt{2}\,\sin\alpha)>0$, 
and it is unbounded from below as a function
of $\overline{p}$.  The NEC is also violated because for any null
vector $t^\alpha$ in the direction perpendicular to ${\bf p}$, one
finds $\langle\psi|\hat{T}_{\alpha\beta}^{\rm ren}
|\psi\rangle t^\alpha t^\beta \approx 
\langle\psi|\hat{T}_{00}^{\rm ren}
|\psi\rangle (t^0)^2$.
As anticipated, the violation of the weak energy condition is associated 
with the 
%
oscillating term in Eq.~(\ref{simpleeq}).  

Our review reflects the view that overcoming the obstacles to forming CTCs and 
wormholes has become increasing difficult. The opposite view is taken in a review with extensive references by Lemos {\sl et al.}~\cite{lemos}.

\subsection{Results and implications}

The NEC can be replaced by the ANEC (or a condition similar to it)
in proving some singularity theorems~\cite{Tipler2,Borde,Roman1} as
well as chronology protection and topological censorship.  It has been
shown that the ANEC holds in free quantum scalar field theory
in Minkowski space and in two-dimensional curved spacetime 
under certain assumptions~\cite{Yurtsever1,Klinkhammer2,wy1}.  However,
Klinkhammer~\cite{Klinkhammer2} 
has pointed out that the ANEC does not hold if one
compactifies any of the space dimensions in Minkowski space.  If, for
example, one identifies the space coordinate $x$ with
$x + L$ in two-dimensional Minkowski space, 
then the energy momentum tensor for the massless scalar field
in the vacuum state is non-zero due to the Casimir effect and given 
by~\cite{DeWitt,df,Kay}
$\langle 0|\hat{T}_{00}\hspace{-1mm}^{\rm ren}|0\rangle =
\langle 0|\hat{T}_{11}^{\rm ren}|0\rangle =
-\pi/6L^2$ and $\langle 0|\hat{T}_{01}^{\rm ren}|0\rangle
= 0$.  Hence for $k^0 = \pm k^1 = 1$, one has
$\langle 0|\hat{T}_{\alpha\beta}^{\rm ren}|0\rangle
k^\alpha k^\beta = -\pi/3L^2 = {\rm const}$, and the ANEC is necessarily
violated.   Wald and Yurtsever also point out that the ANEC
is violated in generic spacetimes for the minimally-coupled massless
scalar field theory~\cite{wy1}. (However, it would be physically wrong to conclude
that the Casimir effect with perfectly reflecting mirrors violates the
ANEC.  See, e.g. \cite{go05}.)

There have been some attempts to rescue the ANEC. Yurtsever proposed
that the integral of
$\langle \psi|T_{\alpha\beta}^{\rm ren}|\psi\rangle k^\alpha k^\beta$
along the null geodesic, to which $k^\alpha$ is tangent, may be bounded
below by a state-independent constant, and he shows that, if true, it can be
used to rule out some wormhole spacetimes~\cite{Yurtsever2}.  Flanagan
and Wald investigated the ANEC smeared over Planck scale
in spacetimes close to Minkowski space, imposing the semi-classical
Einstein equations and showed that it holds for pure and mixed states
if the curvature scale is much larger than the Planck scale and if
incoming gravitational waves do not dominate the 
spacetime curvature~\cite{FlaWald1}.

For a minimally coupled scalar field, the classical energy-density 
$\rho \equiv T_{\alpha\beta}t^\alpha t^\beta$, where $t^\alpha$
is a timelike vector of unit length, is positive definite, and
can be given in the following form:
\be
\rho(x) = \sum_{j}\left[ P^{(j)}\varphi(x)\right]^2\,,
\ee
where $P^{(j)}$ is a differential operator with smooth coefficients.
Let us define the energy-density operator, in the corresponding quantum
theory, normal-ordered 
with respect to a reference state $|\psi_0\rangle$ 
as follows:
\be
:\hspace{-1mm}\hat{\rho}(x)\hspace{-1mm}:
\,\equiv 
\sum_{j}\left\{\left[ P^{(j)}\hat{\varphi}(x)\right]^2 -
\langle\psi_0|\left[P^{(j)}\hat{\varphi}(x)\right]^2
|\psi_0\rangle\right\}\,.  \label{density}
\ee
Although each of these terms is infinite, the difference is
finite and uniquely determined.
This operator differs from the renormalized energy-density operator
by a smooth c-number function which depends only on spacetime
properties. Now, let $x=\gamma(t)$ be a timelike curve and $g(t)$
be a smooth real function satisfying
$\int_{-\infty}^{+\infty} g^2(t)\,dt = 1$.
A QEI takes the following form in general:
\be
\int_{-\infty}^{+\infty}dt\, g^2(t)\langle\psi|\hat{\rho}[\gamma(t)]
|\psi\rangle \geq -C(\gamma,g)
\ee
for any Hadamard state $|\psi\rangle$, where
$C(\gamma,g)$ is a positive number independent of the state
$|\psi\rangle$.

One of the first QEIs was for the minimally-coupled massless scalar
field in four-dimensional Minkowski space with the Lorentzian sampling function.
It has the form~\cite{Ford2,fr1,fr2}
\be
\int_{-\infty}^{+\infty}
dt\,\langle \psi|:\hspace{-1mm}\hat{\rho}(t,{\bf x})
\hspace{-1mm}:|\psi\rangle 
\frac{\tau}{\pi(t^2 + \tau^2)}
\geq -\frac{3}{32\pi^2 \tau^4}\,.  \label{fordineq}
\ee
The QEIs appear to place stringent constraints on spacetimes with 
nontrivial topology, spacetimes with CTCs, as well as on warp-drive~\cite{Alcu} 
spacetimes.  To make these constraints rigorous, however, inequalities in general 
spacetimes are required; and QEIs with compactly-supported sampling functions are 
more useful in these applications.   Flanagan~\cite{flanagan} derived QEIs for
minimally-coupled massless 
scalar field in general two-dimensional curved spacetime with general
smooth sampling functions.  (The bounds given by Flanagan's QEIs are
optimal.)  Fewster was able to establish QEIs in general
globally hyperbolic spacetime in any dimensions using a general
smooth sampling function for minimally-couple scalar field of
arbitrary mass~\cite{fewster1}.  We present his inequality as a
representative of the QEIs.  

Let us consider, for simplicity, a timelike curve given by
${\bf x}=0$ for a given coordinate system in a globally hyperbolic
spacetime and write the minimally-coupled scalar field on this
curve, $\hat{\varphi}(t,{\bf x}=0)$, simply as 
$\hat{\varphi}(t)$.  Fewster's inequality relies 
on the following lemma, which
is an immediate consequence of the work by Radzikowski~\cite{radz1},
which gives the Hadamard condition in the language of microlocal
analysis:
\newtheorem{lemma}{Lemma}
\begin{lemma}\label{lemma1}
Let $P$ be a differential operator with smooth coefficients.
Define the double Fourier transform of the point-separated two-point
function for $P\hat{\varphi}$ 
on a Hadamard state $|\psi_0\rangle$ with a smooth
compactly-supported sampling function $g(t)$ by
\be
\hat{\Delta}_g^P(k_1,k_2) \equiv
\int_{-\infty}^{+\infty} dt_1\int_{-\infty}^{+\infty}dt_2
\, e^{i(-k_1t_1+k_2t_2)}g(t_1)g(t_2)
\langle \psi_0|[P\hat{\varphi}(t_1)] 
[P\hat{\varphi}(t_2)]|\psi_0\rangle\,.
\ee
Then $\hat{\Delta}_g^P(k_1,k_2)$ tends to zero faster than
any polynomial as $k_1\to +\infty$ or $k_2\to +\infty$.
\end{lemma}
The physical interpretation of this lemma is that a Hadamard state
is annihilated by the ``positive-frequency" part
$\int_{-\infty}^{+\infty} dt\,e^{ikt}\hat{\varphi}(t)$ in the limit
$k\to +\infty$. 
The following
result~\cite{fewrev}, a generalization of the argument used
in~\cite{fe}, essentially gives the QEI of Ref.~\cite{fewster1}:
\begin{lemma}
Define the following normal-ordered product:
\be
:\hspace{-1mm}[P\hat{\varphi}(x)]^2\hspace{-1mm}:\,
\equiv [P\hat{\varphi}(t)]^2
- \langle \psi_0|[P\hat{\varphi}(t)]^2|\psi_0\rangle\,.
\ee
Then, the expectation value of 
$:\hspace{-1mm}[P\hat{\varphi}(x)]^2\hspace{-1mm}:$ averaged over
the timelike curve ${\bf x}=0$ with the sampling function $g^2(t)$ 
satisfies
\be
\int_{-\infty}^{+\infty} dt\,g^2(t)
\langle\psi|:\hspace{-1mm}\left[P\hat{\varphi}(t)\right]^2
\hspace{-1mm}:|\psi\rangle
\geq - \frac{1}{\pi}\int_0^\infty\,dk\,
\hat{\Delta}_g^P(k,k)\,.  \label{mainresult}
\ee
The right-hand side is finite by Lemma \ref{lemma1}.
\end{lemma}
\noindent
{\bf Proof}. 
The left-hand side, which we denote by $A$, can be written as
\bea
A & = & \int_{-\infty}^{+\infty} \frac{dk}{2\pi}
\int_{-\infty}^{+\infty} dt_1\int_{-\infty}^{+\infty}dt_2\,g(t_1)
g(t_2)\nonumber\\
&& \times \left\{
\langle\psi|P\varphi(t_1)P\varphi(t_2)|\psi\rangle
- \langle\psi_0|P\varphi(t_1)P\varphi(t_2)|\psi_0\rangle\right\}
e^{-ik(t_1-t_2)}\,.
\eea
The expression inside the curly brackets is invariant under the
interchange $t_1\leftrightarrow t_2$ because
the commutator $[\varphi(t_1),\varphi(t_2)]$ is state-independent.
Hence, one can restrict 
$k$ to be positive and multiply the integral by two.  After this
operation, we see that the first term is of the form
$\int_0^\infty dk\langle\psi|{\cal O}^\dagger(k) {\cal
O}(k)|\psi\rangle \geq 0$ and the second term is exactly the right-hand
side of the inequality (\ref{mainresult}).\\

The Fewster inequality follows immediately from this lemma and
Eq.~(\ref{density}).
\begin{thm}
The expectation value of the normal-ordered energy density
given by Eq.~(\ref{density}) on a Hadamard state $|\psi\rangle$
satisfies the following inequality:
\be
\int_{-\infty}^{+\infty}
dt\, g^2(t)\,\langle\psi|:\hspace{-1mm}
\hat{\rho}(t,{\bf x}=0)\hspace{-1mm}:|\psi\rangle
\geq -\frac{1}{\pi} \sum_{j}
\int_0^\infty dk\,\hat{\Delta}_g^{P^{(j)}}(k,k)\,.
\ee
\end{thm}

Although most of the work on QEIs involves minimally-coupled scalar
fields, corresponding inequalities have also been derived for spin-one fields 
including the electromagnetic field~\cite{pfenning1, fp1} and for spinor 
fields~\cite{vollick,fv1,fm1}.

The QEIs impose severe restrictions on wormhole spacetimes~\cite{fr2}.  Since it is
difficult to calculate the QEI bounds in curved spacetime
(see Ref.~\cite{Pfenning2,ft1} 
for examples of explicit bounds in de Sitter and
other spacetimes), these analyses typically use the QEIs in Minkowski
spacetime, e.g. Eq.~(\ref{fordineq}), with sampling times shorter than
the shortest curvature scale involved in the problem to justify their use.
QEIs are used in this manner to justify the expectation that the
negative energy density, which are necessary for a traversable wormhole
to exist as we have seen, cannot be larger than $C^2\hbar/r_m^4$, where
$r_m$ is the smallest scale in the problem and $C$ is a number which is
typically taken to be $10^2$ or so to err on the conservative side. 
Then the argument to put restrictions on wormholes goes very roughly as
follows.
For a wormhole of size $r_0$ with typical curvature scale
$1/r_0^2$ at the throat, the Einstein equations there tell us that
$G/r_0^2 \lesssim C\hbar/r_m^4$.  Thus, $r_m^2/r_0 \lesssim C\ell_p$,
where $\ell_p = \sqrt{\hbar G} = 1.6\times 10^{-33}\,{\rm cm}$ is the Planck
length.  If the smallest length scale $r_m$ is comparable to the
wormhole size $r_0$, then the wormhole cannot be much larger than the Planck
length.  To circumvent this conclusion, one needs to have $r_m \ll r_0$,
typically by concentrating the negative energy 
in a very narrow region near the throat. Thus, it appears very difficult 
to construct (theoretically) a traversable wormhole, and
that to do so would require fine-tuning of parameters.  (See 
Refs.~\cite{vkd,kuhfittig} for recent wormhole 
models which attempt to evade the argument of Ref.~\cite{fr2}. The 
apparent evasion of the constraints relies on evaluating the energy 
in one frame, and frame-independent statement of the QEIs reinstates 
the constraint~\cite{FewRom1,kuhfittig06}.)

A similar argument was used in Ref.~\cite{ForPfe} to demonstrate
that the warp-drive spacetime as given in Ref.~\cite{Alcu} is
incompatible with ordinary quantum field theory.  This spacetime
contains a bubble of nearly flat spacetime which could move faster than
light. (Olum shows that superluminal travel of this sort, defined
suitably, must involve violation of WEC~\cite{olum1}.)
A spaceship could sit inside this bubble and thus 
move superluminally with respect to the metric outside the bubble.  
However, there is a sphere with a finite thickness where the energy must 
be negative to satisfy the Einstein equations. 
It turns out that the smallest scale involved is the thickness $\Delta$
of the sphere of negative energy.  The situation is then similar to
that of a wormhole with $r_m \sim r_0$.  The negative energy must be
concentrated on a sphere of thickness at most a few orders of magnitude
larger than $\ell_p$, and the energy density there must be only a few
orders of magnitude smaller than $\hbar/\ell_p^4$.  The total negative
energy needed is estimated to be much larger than the (positive) total
mass of the visible universe for a bubble size large enough to fit in
a spaceship.

\subsection{Future problems}

As we have seen, the ANEC is known to be violated even in ordinary
quantum field theory in curved spacetime.  
It would be interesting to see if Yurtsever's
proposal of bounding the ANEC integral from below can be realized in a
useful way.  It is also interesting to find out whether it continues to
be satisfied in spacetimes which are not close to Minkowski space if
one enforces the semi-classical Einstein equations.  It would also be
useful to investigate the validity of the ANEC for interacting
field theories (in Minkowski space to start with).  Interestingly,
Verch has shown that the ANEC holds in two-dimensional Minkowski
spacetime for any interacting field theory with a mass gap~\cite{Verch1}.

QEIs have so far been established primarily for free fields.  Recently, 
Fewster and Hollands~\cite{FewHoll} have shown that there
are QEIs in two-dimensional conformal field theories in Minkowski
space.  It would be interesting to investigate QEIs
in more general interacting field theories, considering the recent
work in which a static negative energy region was constructed using a
2+1 dimensional interacting field theory~\cite{olumgraham}.  One should
also investigate a possible role the QEIs could play in recently
proposed models of dark energy with the negative pressure exceeding the
energy density~\cite{parker,melchiorri,carroll}.

The QEIs place bounds only on the
{\em difference} of the averaged energy density of the given state and
that of the reference state.  Strictly speaking, they have nothing to say
about the averaged energy density of a single state, and they therefore 
do not restrict Casimir-type energy.  The average energy
density of the reference state was implicitly asssumed to be at most of
order $\hbar/r_m^4$, where $r_m$ was the smallest characteristic scale. 
This assumption is reasonable, but attempts to rigorously justify it are 
still in their infancy. It is possible to apply the QEIs to find bounds 
on Casimir-type energies~\cite{fewsterpfenning}, and ``absolute QEIs",
which do not need reference states, have been 
obtained recently~\cite{fewstersmith,smith}.

\acknowledgements{
We thank Chris Fewster, Greg Galloway, Bernard Kay, Ken Olum, Amos Ori and Tom Roman for useful conversations and correspondence.  This work was supported in part by NSF 
Grant 0503366.  
}

\end{document}